\documentclass[twoside,twocolumn,9pt]{article}
\usepackage{extsizes}
\usepackage[super,sort&compress,comma]{natbib} 
\usepackage[version=3]{mhchem}
\usepackage[left=1.5cm, right=1.5cm, top=1.785cm, bottom=2.0cm]{geometry}
\usepackage{balance}
\usepackage{times,mathptmx}
\usepackage{sectsty}
\usepackage{graphicx} 
\usepackage{lastpage}
\usepackage[format=plain,justification=raggedright,singlelinecheck=false,font={stretch=1.125,small,sf},labelfont=bf,labelsep=space]{caption}
\usepackage{float}
\usepackage{fancyhdr}
\usepackage{fnpos}
\usepackage[english]{babel}
\usepackage{array}
\usepackage{droidsans}
\usepackage{charter}
\usepackage[T1]{fontenc}
\usepackage[usenames,dvipsnames]{xcolor}
\usepackage{setspace}
\usepackage[compact]{titlesec}
\usepackage{epstopdf}%This line makes .eps figures into .pdf - please comment out if not required.

\definecolor{cream}{RGB}{222,217,201}

\begin{document}

\pagestyle{fancy}
\thispagestyle{plain}
%\fancypagestyle{plain}{

%%%HEADER%%%
%\fancyhead[C]{\includegraphics[width=18.5cm]{head_foot/header_bar}}
%\fancyhead[L]{\hspace{0cm}\vspace{1.5cm}\includegraphics[height=30pt]{head_foot/journal_name}}
%\fancyhead[R]{\hspace{0cm}\vspace{1.7cm}\includegraphics[height=55pt]{head_foot/RSC_LOGO_CMYK}}
%\renewcommand{\headrulewidth}{0pt}
%}
%%%END OF HEADER%%%

%%%PAGE SETUP - Please do not change any commands within this section%%%
\makeFNbottom
\makeatletter
\renewcommand\LARGE{\@setfontsize\LARGE{15pt}{17}}
\renewcommand\Large{\@setfontsize\Large{12pt}{14}}
\renewcommand\large{\@setfontsize\large{10pt}{12}}
\renewcommand\footnotesize{\@setfontsize\footnotesize{7pt}{10}}
\makeatother

\renewcommand{\thefootnote}{\fnsymbol{footnote}}
\renewcommand\footnoterule{\vspace*{1pt}% 
\color{cream}\hrule width 3.5in height 0.4pt \color{black}\vspace*{5pt}} 
\setcounter{secnumdepth}{5}

\makeatletter 
\renewcommand\@biblabel[1]{#1}            
\renewcommand\@makefntext[1]% 
{\noindent\makebox[0pt][r]{\@thefnmark\,}#1}
\makeatother 
\renewcommand{\figurename}{\small{Fig.}~}
\sectionfont{\sffamily\Large}
\subsectionfont{\normalsize}
\subsubsectionfont{\bf}
\setstretch{1.125} %In particular, please do not alter this line.
\setlength{\skip\footins}{0.8cm}
\setlength{\footnotesep}{0.25cm}
\setlength{\jot}{10pt}
\titlespacing*{\section}{0pt}{4pt}{4pt}
\titlespacing*{\subsection}{0pt}{15pt}{1pt}
%%%END OF PAGE SETUP%%%

%%%FOOTER%%%
%\fancyfoot{}
%\fancyfoot[LO,RE]{\vspace{-7.1pt}\includegraphics[height=9pt]{head_foot/LF}}
%\fancyfoot[CO]{\vspace{-7.1pt}\hspace{13.2cm}\includegraphics{head_foot/RF}}
%\fancyfoot[CE]{\vspace{-7.2pt}\hspace{-14.2cm}\includegraphics{head_foot/RF}}
%\fancyfoot[RO]{\footnotesize{\sffamily{1--\pageref{LastPage} ~\textbar  \hspace{2pt}\thepage}}}
%\fancyfoot[LE]{\footnotesize{\sffamily{\thepage~\textbar\hspace{3.45cm} 1--\pageref{LastPage}}}}
%\fancyhead{}
%\renewcommand{\headrulewidth}{0pt} 
%\renewcommand{\footrulewidth}{0pt}
%\setlength{\arrayrulewidth}{1pt}
%\setlength{\columnsep}{6.5mm}
%\setlength\bibsep{1pt}
%%%END OF FOOTER%%%

%%%FIGURE SETUP - please do not change any commands within this section%%%
\makeatletter 
\newlength{\figrulesep} 
\setlength{\figrulesep}{0.5\textfloatsep} 

\newcommand{\topfigrule}{\vspace*{-1pt}% 
\noindent{\color{cream}\rule[-\figrulesep]{\columnwidth}{1.5pt}} }

\newcommand{\botfigrule}{\vspace*{-2pt}% 
\noindent{\color{cream}\rule[\figrulesep]{\columnwidth}{1.5pt}} }

\newcommand{\dblfigrule}{\vspace*{-1pt}% 
\noindent{\color{cream}\rule[-\figrulesep]{\textwidth}{1.5pt}} }

\makeatother
%%%END OF FIGURE SETUP%%%

%%%TITLE, AUTHORS AND ABSTRACT%%%
\twocolumn[
  \begin{@twocolumnfalse}
%\vspace{3cm}
%\sffamily
%\begin{tabular}{m{4.5cm} p{13.5cm} }

%\includegraphics{head_foot/DOI} 
 \noindent\LARGE{\textbf{The role of membrane curvature for the wrapping of nanoparticles}} \\%Article title goes here instead of the text "This is the title"
%\vspace{0.3cm}  \vspace{0.3cm} \\

  \noindent\large{\bf Amir Houshang Bahrami,\textit{$^{\ddag}$} Reinhard Lipowsky, and Thomas R.\ Weikl} \\%Author names go here instead of "Full name", etc.

%% abstract
%\includegraphics{head_foot/dates} 
 \noindent\normalsize{Cellular internalization of nanoparticles requires the full wrapping of the nanoparticles by the cell membrane. This wrapping process can occur spontaneously if the adhesive interactions between the nanoparticles and the membranes are sufficiently strong to compensate for the cost of membrane bending. In this article, we show that the membrane curvature prior to wrapping plays a key role for the wrapping process, besides the size and shape of the nanoparticles that have been investigated in recent years. For membrane segments that initially bulge away from nanoparticles by having a mean curvature of the same sign as the mean curvature of the particle surface}, we find strongly stable partially wrapped states that can prevent full wrapping. For membrane segments that initially bulge towards the nanoparticles, in contrast, partially wrapped states can constitute a significant energetic barrier for the wrapping process. 

%\end{tabular}

 \end{@twocolumnfalse} \vspace{1.2cm}

  ]
%%%END OF TITLE, AUTHORS AND ABSTRACT%%%

%%%FONT SETUP - please do not change any commands within this section
\renewcommand*\rmdefault{bch}\normalfont\upshape
\rmfamily
\section*{}
\vspace{-1cm}

%%%FOOTNOTES%%%

\footnotetext{Max Planck Institute of Colloids and Interfaces, Department of Theory and Bio-Systems, Science Park Golm, 14424 Potsdam,Germany}

%Please use \dag to cite the ESI in the main text of the article.
%If you article does not have ESI please remove the the \dag symbol from the title and the footnotetext below.
%\footnotetext{\dag~Electronic Supplementary Information (ESI) available: [details of any supplementary information available should be included here]. See DOI: 10.1039/b000000x/}
%additional addresses can be cited as above using the lower-case letters, c, d, e... If all authors are from the same address, no letter is required

\footnotetext{\ddag~Present address: Max Planck Institute of Biophysics, Department of Theoretical Biophysics, Max-von-Laue-Straße 3, 60438 Frankfurt am Main, Germany}

%%%END OF FOOTNOTES%%%

%%%
\section{Introduction}
%%%

Advances in nanotechnology have led to an increasing interest in how nanoparticles interact with living organisms \cite{Nel09}. On the one hand, biomedically designed nanoparticles are promising delivery vehicles in drug treatments \cite{Barbe04,DeJong08,Davis08,Petros10,Wu11b}. On the other hand, the wide application of industrial nanoparticles has also led to safety concerns \cite{Oberdoerster05,Leroueil07,DeJong08}. A current focus is to understand the interactions of nanoparticles with biomembranes that surround the cells and organelles of living organisms. Nanoparticles that are larger than the membrane thickness typically cross membranes by wrapping and subsequent fission of a membrane neck. The wrapping of the nanoparticles can occur spontaneously if the adhesive interactions between the nanoparticles and the membrane are sufficiently strong to compensate for the cost of membrane bending \cite{Lipowsky98,Deserno04,Bahrami14}. The spontaneous wrapping of nanoparticles has been investigated in experiments with lipid vesicles \cite{Dietrich97,Koltover99,Fery03,LeBihan09,Michel12}, polymersomes \cite{Jaskiewicz12,Jaskiewicz12b}, cells \cite{Rothen06,Liu11}, and in theoretical approaches \cite{Lipowsky98,Deserno02,Deserno04,Fleck04,Gozdz07,Benoit07,Nowak08,Decuzzi08,Chen09,Yi11,Cao11,Raatz14,Agudo15}  and simulations  \cite{Noguchi02,Smith07,Fosnaric09,Li10,Yang10,Vacha11,Shi11,Yue11,Bahrami12,Saric12b,Yue12,Dasgupta13,Bahrami13,Dasgupta14,Huang13,Curtis15}.  A recent focus of theoretical investigations and simulations has been on how the size and shape of nanoparticles affects the interplay of adhesion and bending energies during membrane wrapping \cite{Shi11,Vacha11,Li12,Bahrami13,Dasgupta13,Huang13,Yang13,Bahrami14,Dasgupta14,Yi14}. 

\begin{figure}[t]
\centering
\includegraphics[width=\columnwidth]{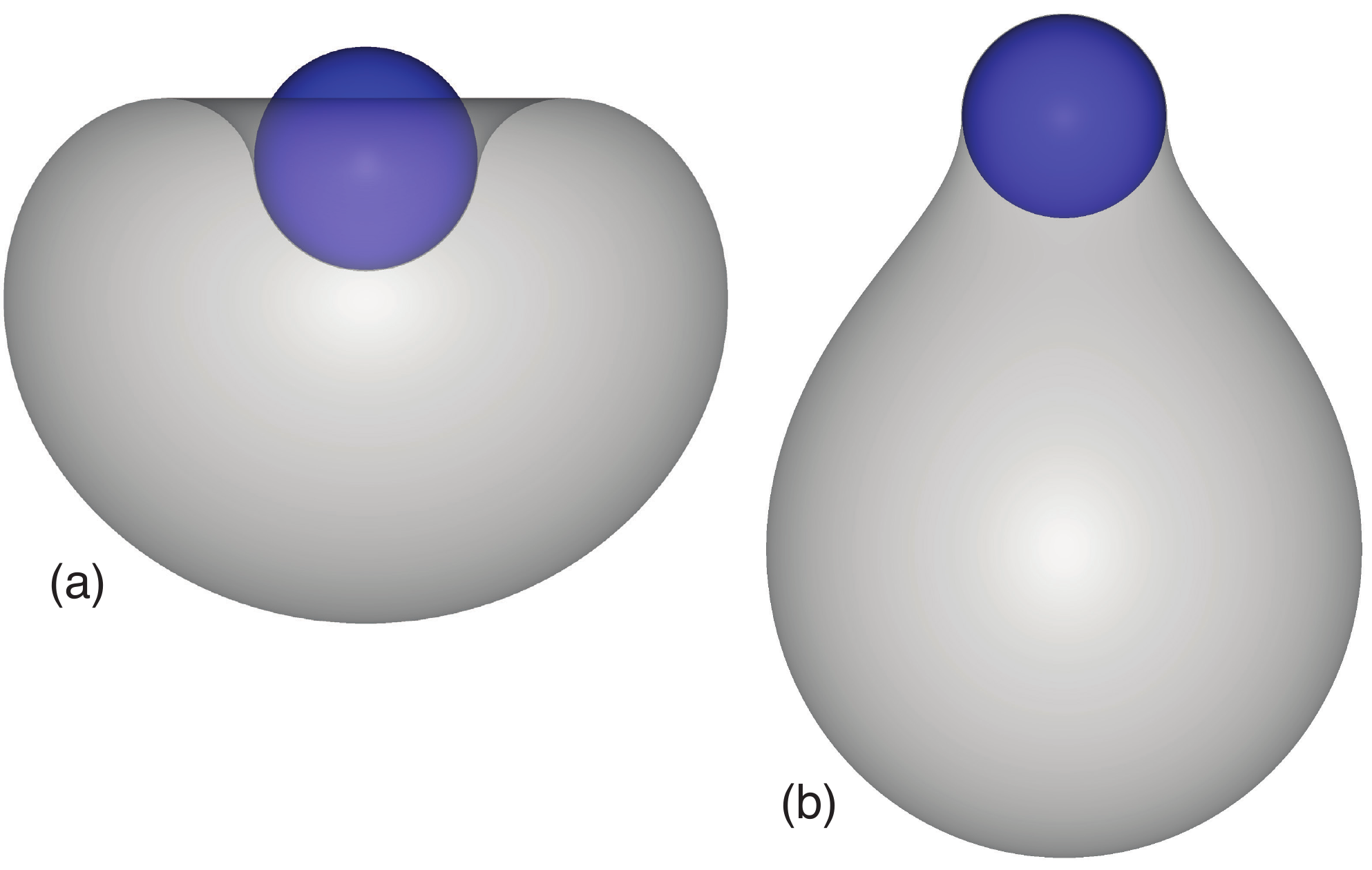}
\caption{A particle (a) outside and (b) inside a vesicle that is half wrapped by the vesicle membrane. For the outside particle, this partially wrapped state is higher in energy than the unwrapped or fully wrapped state at intermediate rescaled adhesion energies $u = W R_p^2/\kappa$ where $W$ is the adhesion free energy per area, $R_p$ is the particle radius, and $\kappa$ is the bending rigidity of the vesicle membrane. For the rescaled adhesion energy $u=2$ at which the unwrapped and fully wrapped state of the particle have equal total energies, the half wrapped state shown in (a) corresponds to an energy barrier of height  $7.0 \kappa$ for the wrapping process. This energy barrier is large since typical values of $\kappa$ are around $20 k_B T$ where $k_B T$ is the thermal energy. For the inside particle, partially wrapped states are lower in energy than the unwrapped state and fully wrapped state at intermediate values of $u$. For $u=2$, the half wrapped state shown in (b) is the minimum-energy state, which is $3.4 \kappa$ lower in energy than both the unwrapped and fully wrapped state and, thus, highly stable for typical values of $\kappa$. The relative curvature here is $c_r = 0.322$ for the outside particle and $c_r = -0.322$ for the inside particle.}
\label{figure_snapshots}
\end{figure}

In this article, we demonstrate that the wrapping of nanoparticles by membranes is also affected by the membrane curvature before wrapping, i.e.~on whether the membrane initially bulges towards or away from a particle.  As a model system, we consider a spherical particle that is initially either inside or outside of a membrane vesicle. The vesicle considered here can freely adjust its volume and adopts an overall spherical shape that is locally distorted by the wrapping of the nanoparticle. Before wrapping, the vesicle bulges towards an outside particle, and away from an inside particle.  The equilibrium behavior of such a vesicle and a spherical particle, as obtained by minimization of the combined bending and adhesion free energy of the vesicle membrane, can be described by two dimensionless parameters, the relative curvature $c_r=\pm  R_p/R_v$ of the particle with radius $R_p$ and the vesicle with initial radius $R_v$, and the rescaled adhesion energy $u$. As sign convention, we choose $c_r = + R_p/R_v$ for outside particles and $c_r = -R_p/R_v$ for inside particles. This two-dimensional parameter space is a subset of the general four-dimensionless parameter space, which includes the spontaneous curvature and the osmotic pressure difference across the vesicle membrane as additional parameters \cite{Agudo15}. This four-dimensional parameter space is divided up into four distinct engulfment or wrapping regimes corresponding to no wrapping, full wrapping, partial wrapping, as well as a bistable  regime with the coexistence of unwrapped and fully wrapped states \cite{Agudo15}. 

For vanishing spontaneous curvature and vanishing osmotic pressure difference as considered here, the wrapping scenario is dominated by the relative curvature $c_r$. We demonstrate that a particle that is initially outside of the vesicle is either not wrapped or fully wrapped by the vesicle membrane. Partially wrapped states of such an outside particle with positive relative curvature $c_r$ have a higher energy than the unwrapped or fully wrapped state and, thus, constitute an energy barrier for the wrapping process. A particle inside the vesicle with negative $c_r$, in contrast, can exhibit stable partially wrapped states that are significantly lower in energy than the unwrapped or fully wrapped state (see Fig.\ \ref{figure_snapshots}). Our results thus indicate different wrapping scenarios that depend on whether the membrane initially bulges away or towards a nanoparticle, in agreement with the stability analysis in Ref.\  31.

Our results have direct implications for the spontaneous internalization of spherical nanoparticles by cells. For segments of cellular membranes that initially bulge away from the nanoparticles, we find that the full wrapping required for internalization can be impeded by highly stable partially wrapped states. The particles adhere to the membranes in these partially wrapped states, but are not internalized. For membrane segments that initially bulge towards the nanoparticles, partially wrapped states constitute an energetic barrier for full wrapping and internalization. 

\begin{figure*}[t]
\centering
\includegraphics[width=2\columnwidth]{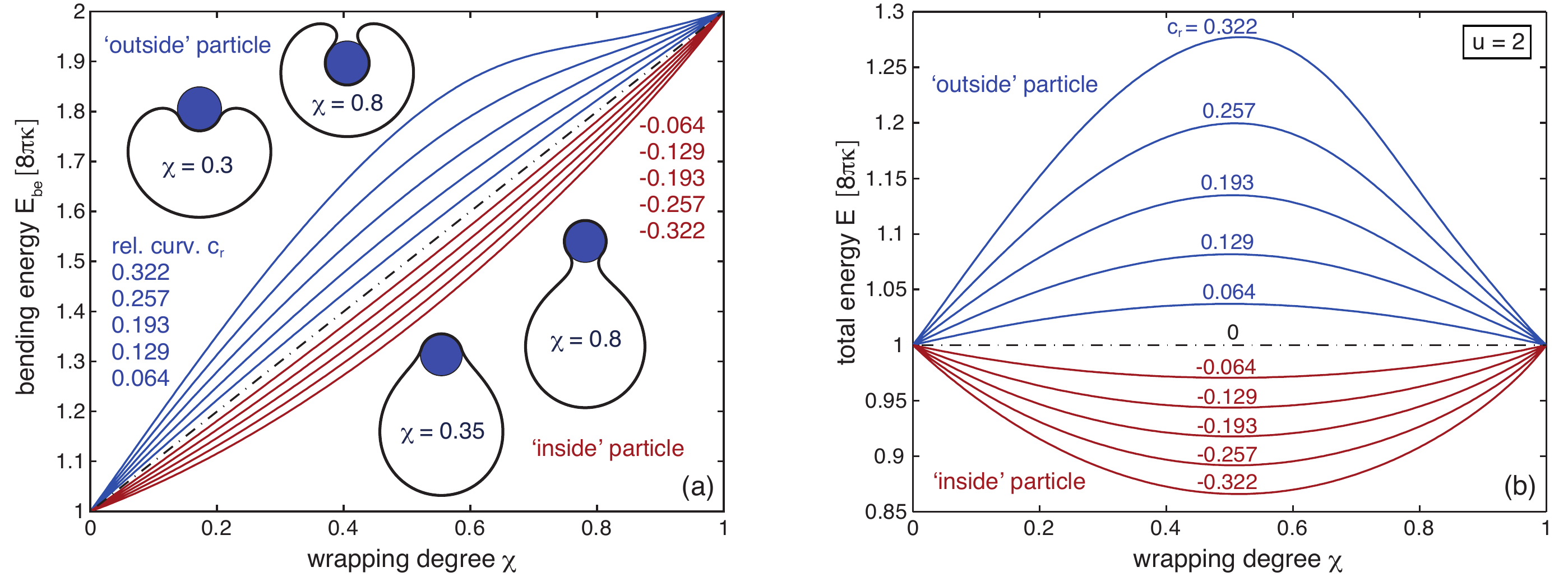}
\caption{(a) Bending energy $E_{be}$ of a vesicle wrapping a spherical nanoparticle as function of the wrapping degree $\chi$ for different relative curvatures $c_r=\pm R_p/R_v$ of the particle with radius $R_p$ and vesicle with initial radius $R_v$ at $\chi=0$. 
The relative curvature $c_r$ is positive for a particle outside the vesicle (blue curves) and negative for a particle inside the vesicle (red curves). The black dotted curve corresponds to a large vesicle or a flat bilayer with $c_r=0$. The relative curvatures for the shown vesicle shapes are $c_r = 0.322$ (top) and $c_r = -0.322$ (bottom).
-- (b) Total energy $E$ of a vesicle wrapping a spherical nanoparticle as a function of the wrapping degree $\chi$ for the rescaled adhesion energy $u=2$. At this rescaled adhesion energy, the unwrapped state with $\chi=0$ and the fully wrapped state with $\chi=1$ have equal energies $E$.  }
\label{figure_energies}
\end{figure*}
\section{Model}

The spontaneous wrapping of nanoparticles by membranes is governed by the interplay of the adhesion energy of the particles and the bending energies of the membranes \cite{Lipowsky98,Bahrami14}. The bending energy of a vesicle membrane with negligible spontaneous curvature is the integral \cite{Helfrich73}
\begin{equation}
  E_{be} = 2\kappa \oint M^{2} dA 
 \label{ebend}
\end{equation}
over the total surface $A$. Here, $M$ is the local mean curvature of the membrane, and $\kappa$ is the membrane's bending rigidity. The shape of an axisymmetric vesicle considered here can be described by the function $\psi(s)$ where $\psi$ is the angle between the tangent of the vesicle contour and the direction $r$ perpendicular to the rotational symmetry axis, and $s$ is the arc length 
\cite{Seifert91}. 
In this parametrization, the principal curvatures can be written as $c_1=\dot{\psi}=d\psi/ds$ and $c_2=\sin\,\psi/r$, and the local  mean curvature is $M= \frac{1}{2}(c_1+c_2)=(\dot{\psi}+\sin\,\psi/r)/2$. The dots here denote derivatives with respect to the arc length $s$. The radial distance $r$ from the rotational symmetry axis is related to the angle $\psi$ via the constraint $\dot{r}=\cos\,\psi$ \cite{Seifert91}. The total area $A$ of the vesicle is the sum $A=A_u+A_b$ of the area $A_b=4\pi R_p^2 \chi$ of the bound membrane segment  and the area  $A_u$ of the unbound segment. Here, $\chi$ is the wrapping degree of the particle, i.e.~the fraction of the particle's area that is covered by the membrane. The total energy functional then has the form $F=E_{be}+\gamma(s)(\dot{r}-\cos\,\psi)+\sigma A$ with Lagrange multipliers $\gamma(s)$ and $\sigma$ that account for the constraints on the radial coordinate and the membrane area. The volume of the vesicle can freely adapt, which corresponds to an osmotic pressure difference $\Delta P = 0$ accross the vesicle membrane. Minimization of the energy functional $F$ for the unbound segment of the vesicle leads to the Euler-Lagrange equations \cite{Seifert91}
\begin{equation}
\ddot{\psi}= \frac{\cos\,\psi}{r}\left(\frac{\sin\,\psi}{r}-\dot{\psi}\right) + \frac{\gamma}{r}\sin\,\psi, \label{saieq}
\end{equation}
\begin{equation}
\dot{\gamma}=\frac{1}{2}\left(\dot{\psi}^2-\frac{\sin^2\,\psi}{r^2}\right)+\sigma   \label{gamaeq}
\end{equation}
We determine numerical solutions of these equations with the Runge-Kutta algorithm \cite{Seifert91} for different wrapping degrees $\chi$ of the particle. A solution $\psi(s)$
describes the conformation of the unbound segment of the vesicle with minimum bending energy $E_{be}$ for a given value of the wrapping degree $\chi$ 
of the nanoparticle. 

For a given wrapping degree $\chi$ of the particle, the total energy is the sum
\begin{equation}
E(\chi)=E_{be}(\chi)+ E_{ad}(\chi)=E_{be}(\chi)- W A_p  \chi
\label{total_energy}
\end{equation}
of the adhesion energy $E_{ad}(\chi)$ of the spherical particle with surface area $A_p = 4 \pi R_p^2$ and the minimum bending $E_{be}(\chi)$ of the vesicle at this wrapping degree. Besides $\chi$, the total energy (\ref{total_energy}) depends on the adhesion free energy per area $W$ \cite{Seifert90}, the bending rigidity $\kappa$, the radius $R_p$ of the spherical particle, and the initial radius $R_v$ of the spherical vesicle prior to wrapping. Since we are free to choose an energy scale and a length scale as units in our model, these four parameters can be reduced to two independent, dimensionless parameters. We choose here the rescaled adhesion energy $u= W R_p^2 /\kappa$ and the relative curvature $c_r = \pm  R_p/R_v$ of the vesicle with initial curvature $1/R_v$ and the nanoparticle with curvature $1/R_p$. To distinguish particles inside and outside the vesicle, we use the sign convention $c_r = + R_p/R_v$ for outside particles and $c_r = -R_p/R_v$ for inside particles.

%%%
\section{Results}
%%%

The minimum bending energy $E_{be}(\chi)$ of a vesicle that wraps a particle to a degree $\chi$ depends on whether the particle is located inside or outside the vesicle (see Fig.\ \ref{figure_energies}(a)). At intermediate wrapping degrees $0< \chi < 1$, the bending $E_{be}(\chi)$ of a vesicle wrapping an outside particle is larger than the bending energy of vesicle that wraps an inside particle to the same degree $\chi$. For outside particles, the bending energy $E_{be}(\chi)$ increases with the relative curvature $c_r$ of the vesicle and particle (see blue lines in Fig.\ \ref{figure_energies}(a)). For inside particles, in contrast, the bending energy $E_{be}(\chi)$ decreases with an increasing absolute value of  $c_r$ (see red lines in Fig.\ \ref{figure_energies}(a)). At $\chi=0$, all lines in Fig.\ \ref{figure_energies}(a) coalesce into the bending energy $8\pi\kappa$ of a spherical vesicle. The spherical shape minimizes the bending energy of a vesicle that can freely adjust its volume \cite{Deuling76}. At $\chi=1$, all lines coalesce into the bending energy $16\pi\kappa$ of two spheres that are connected by an infinitesimally small catenoidal neck of zero mean curvature $M$ and, thus, zero bending energy.  The smaller one of these two spheres completely wraps the particle. 

Since the adhesion energy $E_{ad}(\chi) = - 4 \pi \kappa u \chi$ is proportional to the wrapping degree $\chi$, the total energy $E(\chi) =  E_{be}(\chi)- 4 \pi \kappa u \chi$ for different rescaled adhesion energies $u$ can be easily obtained by `tilting' the curves $E_{be}(\chi)$ shown in Fig.\ \ref{figure_energies}(a). The total energy $E(\chi)$ at the rescaled adhesion energy $u=2$ is displayed in Fig.\ \ref{figure_energies}(b). For $u=2$, the adhesion energy $E_{ad}$ for the fully wrapped state of the particle is equal to $-8\pi\kappa$. The fully wrapped state with $\chi=1$ and the unwrapped state with $\chi=0$ therefore have the same total energies at the value $u=2$ of the rescaled adhesion energy \cite{Lipowsky98,Deserno04}. For an outside particle, the total energy of partially wrapped states with $0<\chi<1$ is higher than the total energy of the fully wrapped and unwrapped states (see blue lines in Fig.\ \ref{figure_energies}(b)). The total energy is maximal at half wrapping with $\chi=0.5$ and strongly increases with the relative curvature $c_r=R_p/R_v$ of the vesicle and the particle. For an inside particle, in contrast, the total energy of partially wrapped states is lower than the total energy of the fully wrapped and unwrapped states (see red lines in Fig.\ \ref{figure_energies}(b)), and minimal at half wrapping.

\begin{figure}[t]
\centering
\includegraphics[width=0.95\columnwidth]{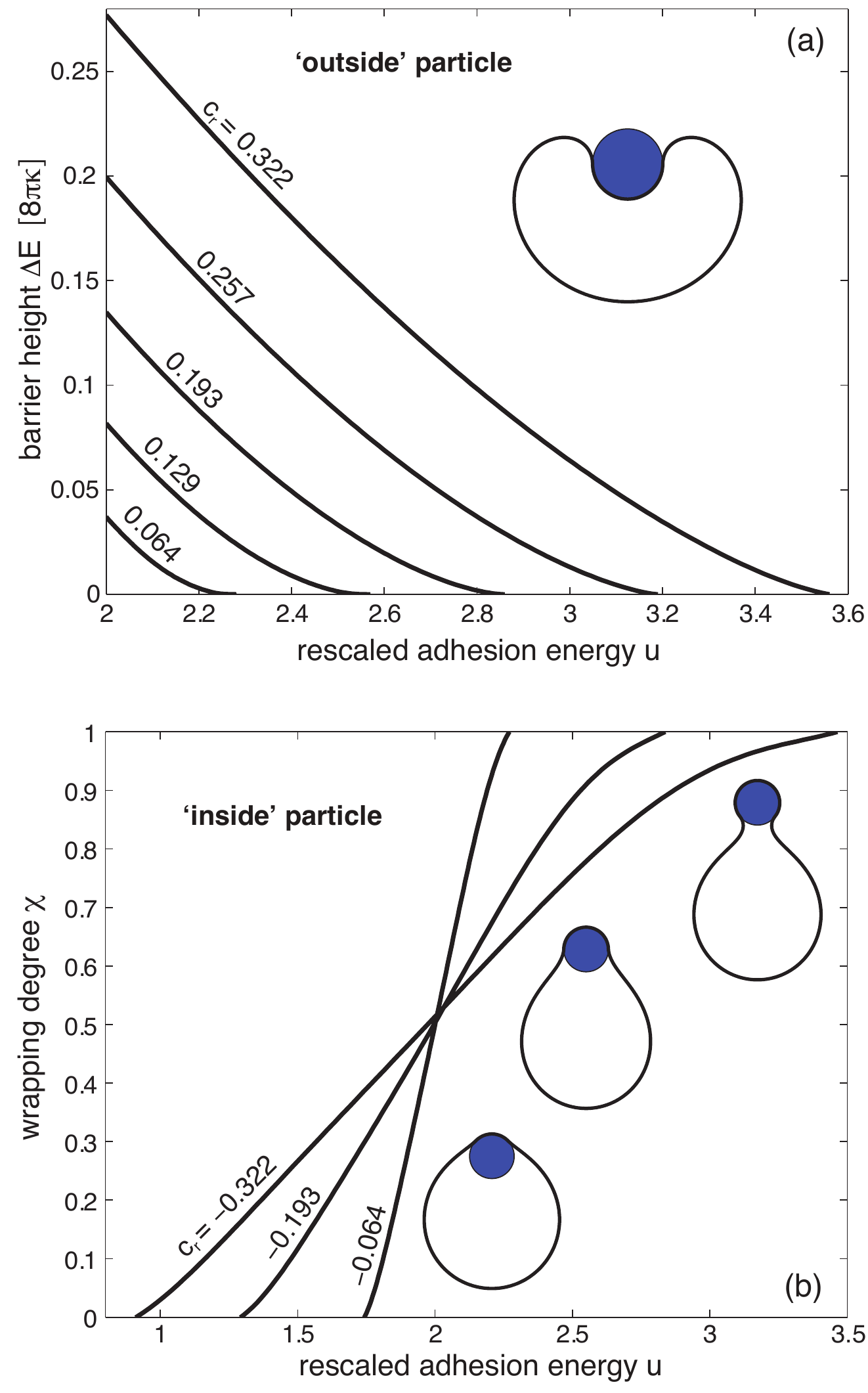}
\caption{(a) Energy barrier $\Delta E$ as a function of the rescaled adhesion energy $u$ for outside particles with different positive relative curvatures $c_r$. The energy barrier $\Delta E$ is the difference between the maximal total energy of partial wrapping and the total energy of the unwrapped state. -- (b) Wrapping degree $\chi$ at which the total energy of an inside particle is minimal as a function of the rescaled adhesion energy $u$.
}
\label{figure_u}
\end{figure}

Partially wrapped states of an outside particle thus constitute an energetic barrier for wrapping at intermediate values of $u$. With increasing rescaled adhesion energy $u$, the barrier decreases and finally vanishes as the total energy curves $E(\chi)$ are `tilting' further and further towards the fully wrapped state (see Fig.\ \ref{figure_u}(a)). In contrast, partially wrapped states of an inside particle are globally stable for a range of intermediate values of $u$ centered around $u=2$. Within this range, the wrapping degree $\xi$ of the inside particle continuously increases with $u$ from $\chi=0$ to $1$ (see Fig.\ \ref{figure_u}(b)).

\begin{figure}[t]
\centering
\includegraphics[width=0.95\columnwidth]{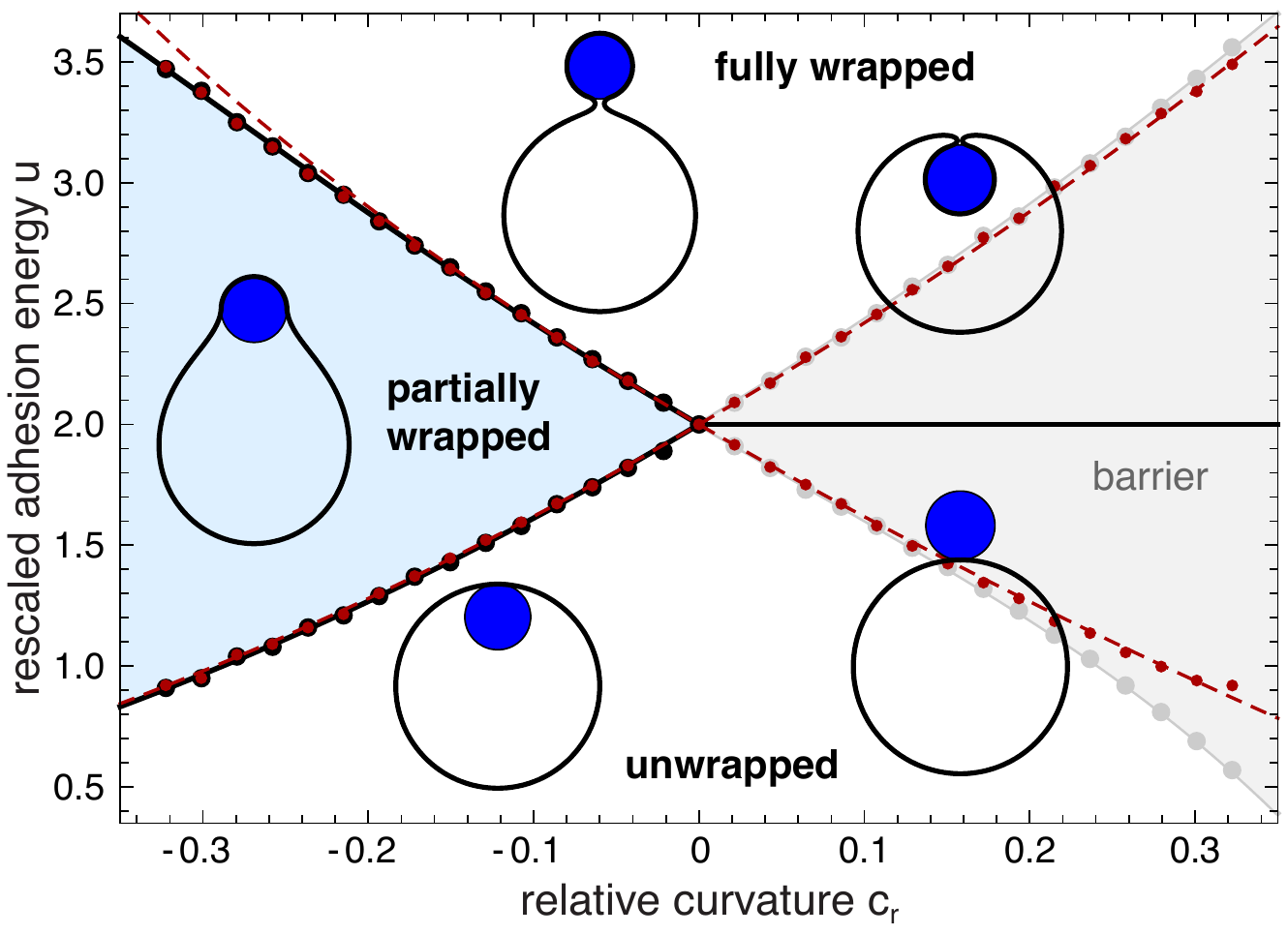}
\caption{Morphology diagram of stable states of a particle adhering to a vesicle. The three black lines divide the diagram into three regions in which the particle is either partially wrapped, unwrapped, or fully wrapped. In the grey shaded region, the transitions between the unwrapped state and the wrapped state of particle outside the vesicle require the crossing of an energy barrier. The upward line through the black and grey data points represents the polynomial fit $u = 2 + 4.1 c_r + 2.2 c_r^2$, and the downward line through the data points represents the fit $u = 2 -4.0 c_r + 1.3 c_r^2 - 4.7 c_r^3 - 10.9 c_r^4$.  The red dashed lines are the instability lines of the unwrapped and fully wrapped state given by the Eqs.\ (\ref{instab_unwrapped}) and  (\ref{instab_wrapped}), which are derived in the appendix from the stability relations in Ref.\ 31. The red points indicate the instabilities derived from our numerical results for the bending energy $E_{be}(\chi)$ (see Fig.\ \ref{figure_energies}(a)). The unwrapped state becomes locally unstable for $\left({\rm d}E(\chi)/{\rm d}\chi\right) |_{\chi=0} = 0$ where $E(\chi) = E_{be}(\chi) - 4 \pi\kappa u \chi$ is the total energy as a function of the wrapping degree $\chi$. The wrapped state becomes locally unstable for $\left({\rm d}E(\chi)/{\rm d}\chi\right) |_{\chi=1} = 0$. For outside particles with large relative curvature $c_r$, the instability lines slightly deviate from the lines at which the barrier vanishes because the derivative of the function $E_{be}(\chi)$ changes its sign both for small and large wrapping degrees $\chi$ close 0 and 1, which leads to secondary satellite minima of the total energy $E(\chi)$ for values of $u$ close to the grey lines at which the barrier vanishes.
}
\label{figure_morphologies}
\end{figure}

The morphology diagram shown in Fig.\ \ref{figure_morphologies} illustrates how the stable wrapping state of a particle depends on the relative curvature $c_r$ and rescaled adhesion energy $u$. The three thick black lines divide the diagram into three regions in which the 
particle is either fully wrapped, partially wrapped, or unwrapped in its equilibrium state. Partially wrapped states (blue shaded region) only occur for negative relative curvatures $c_r$ for which the membrane initially bulges away from the particle inside the vesicle. For positive relative curvatures $c_r$ for which the membrane initially bulges towards the particle outside the vesicle, the fully wrapped state is the equilibrium state for $u>2$, and the unwrapped state is the equilibrium state for $u<2$. Stable partially wrapped states do not occur for positive $c_r$. In the grey shaded region, the wrapping and unwrapping of the particle requires the crossing of an energy barrier. The red dashed lines represent the local instability lines of the unwrapped and fully wrapped states derived from analytical results of Ref.\ 31. The red points indicate the local instabilities of these states determined from our numerical results.

%

%%%
\section{Discussion and conclusions}
%%%

The numerical results presented in this article illustrate that the wrapping of nanoparticles depends on the initial membrane curvature prior to wrapping. Stable partially wrapped states of nanoparticles can occur if the membrane initially bulges away from the particles. In contrast, large energetic barriers for the wrapping process can occur if the membrane initially bulges towards the particles. In the present study, we considered spherical particles and described their interactions with the membrane by a single parameter, the adhesion free energy per area $W$. Partially wrapped states can also occur if the membrane-particle interactions are described by  interaction potentials with  potential ranges that exceed a few percent of the particle radius  \cite{Raatz14}, for non-spherical particles \cite{Dasgupta13,Bahrami13,Bahrami14,Dasgupta14}, and in situations in which the wrapping of particles is opposed by membrane tension \cite{Deserno04,Dietrich97} or by volume constraints and osmotic pressure differences of vesicles \cite{Bahrami12,Agudo15}.

To investigate the local interplay of membrane curvature and nanoparticle adhesion during wrapping, we have focused here on a simple model system in which a vesicle with freely adjustable volume interacts with a spherical particle. The volume of a lipid vesicle is freely adjustable in the absence of osmotically active molecules such as salts or sugars, because water can cross the lipid membrane. Prior to wrapping, the vesicle then adopts a spherical shape with bending energy $8\pi\kappa$ because this spherical shape minimizes the bending energy of the vesicle. During the wrapping process of a particle, the spherical vesicle shape is locally distorted in the vicinity of the particle. In the fully wrapped state of the particle, the vesicle shape consists of two spheres with total bending $16\pi\kappa$, which are connected by a small catenoidal neck of zero bending energy for vanishing spontaneous curvature (see e.g.\ Ref.\ 11 for a more detailed discussion). The smaller one of these spherical vesicle segments fully encloses the particle with an adhesion energy $-4\pi R_p^2U$. For this system, the local interplay of bending and adhesion energies leads to an energy landscape in which the unwrapped and fully wrapped state have equal overall energies at the rescaled adhesion energy $u = U R_p^2/\kappa = 2$, irrespective of the relative size of the vesicle and nanoparticle. 

For a vesicle with constrained volume, in contrast, the wrapping of nanoparticles leads to global changes of the vesicle shape  \cite{Gozdz07,Cao11,Bahrami14,Agudo15}. Volume constraints of a vesicle arise in the presence of osmotically active particles inside and outside the vesicle, because the vesicle then `adjusts' its volume $V$ in such a way that the interior osmotic pressure balances the exterior one. Prior to wrapping, the shape of such a vesicle is determined by its area-to-volume ratio \cite{Deuling76,Seifert91}, as described by the reduced volume $v = 6\sqrt{\pi}\,V/A^{3/2} \le 1$. The maximal value $v =1$ of the reduced volume corresponds to the area-to-volume ratio of a sphere.  During wrapping of a particle, a vesicle with constant area $A$ and volume $V$ changes its shape and becomes `more spherical', because the membrane wrapped around the particle effectively `decreases' the area of the vesicle membrane, and the volume of the particle inside the vesicle `increases' the volume of the vesicle \cite{Bahrami14}. In the fully wrapped state, the overall shape of the vesicle consists of a spherical vesicle segment that fully encloses the nanoparticle, and an unbound vesicle segment with effective reduced volume $v_\text{ef} = 6\sqrt{\pi}(V + V_p)/(A-A_p)^{3/2}$ where $V_p$ and $A_p$ are the volume and surface area of the particle. This effective reduced volume $v_\text{ef}$ is larger than the initial reduced volume $v$ of the vesicle. Because the bending energy of a vesicle decreases with increasing reduced volume, the global shape change of the vesicle facilitates the wrapping of the particle.  This is reflected in the fact that the unwrapped and fully wrapped state of the particle have equal overall energies at rescaled adhesion energies $u<2$, depending on the relative size of the vesicle and particle  and the initial reduced volume of the vesicle \cite{Bahrami14}. The full wrapping of the particle is only possible if the initial reduced volume $v$ of the vesicle is sufficiently small. 

We have considered here the spontaneous wrapping of nanoparticles by vesicle membranes that arises from the interplay of adhesion and bending energies during wrapping. This spontaneous wrapping  has been investigated in experiments with lipid vesicles \cite{Dietrich97,Koltover99,Fery03,LeBihan09,Michel12} and polymersomes \cite{Jaskiewicz12,Jaskiewicz12b}.
The wrapping of nanoparticles by cell membranes can either occur spontaneously \cite{Rothen06,Liu11}, 
or can be driven by the curvature-inducing proteins and protein machineries  involved in endocytosis and phagocytosis \cite{Mukherjee97,Conner03,Hurley10,Rodriguez13}.  While the diameter of cells is of the order of micrometers, cell membranes are often rather strongly curved, e.g.\ from  protrusions such as microvilli that are induced by actin polymerization and have diameters of the order of 100 nm \cite{McConnell09}. Rather strong curvatures also occur in membranes surrounding cellular organelles such as the endoplasmic reticulum \cite{Shibata10}. The effect of the initial membrane curvature on particle wrapping considered in this article may be relevant for the internalization of nanoparticles by cells and cellular organelles in such rather strongly curved membrane regions, even for particles with diameters of the order of 10 nm. For such small nanoparticles, the absolute value $|c_r|$ of the relative curvature can exceed 0.1 in strongly curved regions of cellular membranes, which may lead to stable partially wrapped states or high energetic barriers, depending on the sign of $c_r$ (see Figs.\ \ref{figure_energies} to \ref{figure_morphologies}). A more detailed modeling of the uptake of nanoparticles by cell membranes needs to take into account the spontaneous curvature and increased bending rigidity of membrane segments with a protein coat or scaffold, and the contribution of receptor-mediated adhesion to the adhesion energy $U$ between ligand-coated particles and cell membranes \cite{Agudo15}. Such a detailed modeling has been recently shown to explain the experimentally observed dependence of clathrin-mediated endocytosis on the particle size \cite{Agudo15}. 

%%%
\section*{Appendix}
%%%

\subsection*{Instability of the unwrapped state}

The free, unwrapped state of the particle  is unstable if the initial wrapping by the vesicle membrane leads to a gain in adhesion energy that overcompensates 
the  increase in the membrane's bending energy. The free state becomes unstable along the line \cite{Agudo15}
\begin{equation}
R_p  =  \frac{1}{R_W^{-1} \pm M}
\label{Rp}
\end{equation}
with 
\begin{equation} 
R_W = \sqrt{ 2\kappa/W }   
\end{equation}
where $M$ is the mean curvature of the membrane at the location of the particle prior to wrapping, and the sign in front of $M$ indicates whether the particle is initially inside ($+$) or outside ($-$) the vesicle. Eq.~(\ref{Rp}) holds for $M<1/R_W$ in the case of an outside particle and for $M>-1/R_W$ in the case of an inside particle to ensure that the membrane does not intersect the particle before contact. The vesicle with adjustable volume has a spherical shape with mean curvature 
\begin{equation}
M  = 1/R_v
\end{equation} 
before wrapping. The instability relation (\ref{Rp}) is then 
\begin{equation}
R_p = \frac{1}{R_W^{-1} \pm R_v^{-1}}
\end{equation}
When expressed in terms  of the dimensionless parameters $u=U R_p^2/\kappa$ and $c_r=\pm R_p/R_v$ with $c_r = + R_p/R_v$ for an outside particle and $c_r = -R_p/R_v$ for an inside particle, this relation for the instability of the free, unwrapped state becomes
\begin{equation}
u = 2 (1+c_r)^2 
\label{instab_unwrapped}
\end{equation}

\subsection*{Instability of the fully wrapped state}

In the fully wrapped state, the nanoparticle is fully covered by the membrane, but still connected to the mother vesicle by a  small membrane neck. In the coarse-grained description used here, the fully wrapped state corresponds to a limit shape with an ideal neck that is attached to the  mother vesicle in a single contact point. At this contact point, the unbound membrane segment of the mother vesicle at the location of the nanoparticle has the mean curvature $M^\prime$. The fully wrapped state becomes unstable along the line \cite{Agudo15}

\begin{equation}
R_p =  \frac{1}{R_W^{-1} \pm (M^\prime - 2m)}
\label{Rp2}
\end{equation}
where $m$ is the spontaneous curvature of the membrane, and the sign in front of $M^\prime - 2m$ indicates whether the particle is initially outside ($+$) or inside ($-$) the vesicle.
Eq.~(\ref{Rp2}) holds for $2m - 1/R_W<M^\prime<1/R_p$ in the case of an outside particle, and for $-1/R_p< M^\prime < 2m +1/R_W$  in the case of an inside particle. The relations $M^\prime<1/R_p$ for an outside particle and $-1/R_p< M^\prime$ for an inside particle ensure that the membrane does not intersect the particle. If the mean curvature $M^\prime$ is smaller than  $2m - 1/R_W$ in the case of an outside particle, or larger than $2m +1/R_W$ in the case of an inside particle, the fully wrapped state is unstable for all particle sizes \cite{Agudo15}. 

For a vesicle with adjustable volume, the unbound vesicle membrane has a spherical shape with mean curvature 
\begin{equation}
M^\prime  = 1/\sqrt{R_v^2 - R_p^2} 
\end{equation}
because the overall area of the vesicle membrane is constant. For spontaneous curvature $m = 0$, the instability relation (\ref{Rp2}) is then 
\begin{equation}
R_p = \frac{1}{R_W^{-1} \pm (R_v^2 - R_p^2)^{-1/2}  }
\end{equation}
When expressed in terms  of the  parameters $u$ and $c_r$ used in this article, this relation for the instability of the fully wrapped state becomes
\begin{equation}
u = 2 \left(1 - \frac{c_r}{\sqrt{1 - c_r^2}} \right)^2 
\label{instab_wrapped}
\end{equation}
\section*{Acknowledgements}
Financial support from the Deutsche Forschungsgemeinschaft (DFG) {\em via} the International Research Training Group 1524 ``Self-Assembled Soft Matter Nano-Structures at Interfaces" is gratefully acknowledged.

%%%REFERENCES%%%
%\bibliography{membranes}
%\bibliographystyle{rsc}

\thispagestyle{plain}

\footnotesize
\providecommand*{\mcitethebibliography}{\thebibliography}
\csname @ifundefined\endcsname{endmcitethebibliography}
{\let\endmcitethebibliography\endthebibliography}{}


%\balance


\end{document}